\documentclass[]{mn2e}
\usepackage{graphicx}
\usepackage{epsfig}
\usepackage{amsmath}
\newcommand       \Angstrom     {\,{\rm \AA}}

\newcommand       \cm           {\,{\rm cm}}

\newcommand       \eV           {\,{\rm eV}}

\newcommand       \K            {\,{\rm K}}

\newcommand       \s            {\,{\rm s}}
\newcommand       \sr           {\,{\rm sr}}

\newcommand       \GHz     {\,{\rm GHz}}

\newcommand     \gtsim  {\lower.5ex\hbox{$\buildrel > \over \sim$}}
\newcommand     \ltsim  {\lower.5ex\hbox{$\buildrel < \over \sim$}}
\newcommand     \simgt  {\lower.5ex\hbox{$\buildrel > \over \sim$}}
\newcommand     \simlt  {\lower.5ex\hbox{$\buildrel < \over \sim$}}

\newcommand       \mum          {\,{\rm \mu m}}

\newcommand       \Teff         {T_{\rm eff}}
\newcommand       \Tstar      {T_{\rm eff}}

\newcommand       \simali       {\sim\,}
\newcommand       \magni        {\,{\rm mag}}

\newcommand	  \NH          {N_{\rm H}}
\newcommand           \Cind       {\left[\frac{\rm C}{\rm H}\right]_{\rm ind}}
\newcommand	  \hnuabs    {\langle h\nu \rangle_{\rm abs}}
\def    \Nb	{M}
\def    \bT	{{\bf T}}

\def    \abs       {{\rm abs}}

\title{
Infrared Emission of Specific Polycyclic Aromatic
Hydrocarbon Molecules: Indene
}
\author[Li, Li, Yang \& Fang]
{Kaijun~Li$^{1}$, Aigen Li$^{2}$\thanks{lia@missouri.edu},
              X.J.~Yang$^{3}$\thanks{xjyang@xtu.edu.cn},
             and Taotao~Fang$^{1}$\thanks{fangt@xmu.edu.cn}\\
$^1$Department of Astronomy,
       Xiamen University, Xiamen, Fujian 361005, China\\
$^2$Department of Physics and Astronomy,
                  University of Missouri,
                  Columbia, MO 65211, USA\\
$^3$Hunan Key Laboratory for Stellar
              and Interstellar Physics
              and School of Physics and Optoelectronics,
              Xiangtan University, Hunan 411105, China\\
              }

\begin{document}

\date{}
\pagerange{\pageref{firstpage}--\pageref{lastpage}} \pubyear{2023}

\maketitle

\label{firstpage}
\begin{abstract}
Polycyclic aromatic hydrocarbon (PAH) molecules
have long been suggested to be present
in the interstellar medium (ISM).
Nevertheless, despite their expected ubiquity
and sustained searching efforts,
identifying specific interstellar PAH molecules
from their infrared (IR) spectroscopy
has so far been unsuccessful.
However, due to its unprecedented sensitivity,
the advent of the {\it James Webb Space Telescope}
(JWST) may change this. 
Meanwhile, recent years have witnessed
breakthroughs in detecting specific PAH
molecules (e.g., indene, cyanoindene,
and cyanonaphthalene) through their
rotational lines in the radio frequencies.
%
%
As JWST holds great promise for identifying 
specific PAH molecules in the ISM
based on their vibrational spectra in the IR,
in this work we model the vibrational
excitation of indene,
a molecule composed of
a six-membered benzene ring
fused with a five-membered cyclopentene ring,
and calculate its IR emission spectra
for a number of representative astrophysical regions.
This will facilitate JWST to search for and identify
indene in space through its vibrational bands
and to quantitatively determine or place
an upper limit on its abundance.
\end{abstract}


\section{Introduction}\label{sec:intro}
Polycyclic aromatic hydrocarbon (PAH) molecules
are composed of multiple hexagonal benzene rings.
These aromatic rings join together in flat sheets
whose edges are decorated with hydrogen atoms.
On Earth, PAHs are a common by-product of
the incomplete combustion of organic matter,
and are often seen in auto exhaust and grilled meat.
PAHs are believed to be also widespread and abundant
in extraterrestrial environments (see Li 2020).
The ``unidentified infrared'' emission (UIE) bands  
at 3.3, 6.2, 7.7, 8.6, 11.3 and 12.7$\mum$,\footnote{%
    There is strong consensus
    within the astronomical community
    to name these ``UIE'' bands
    the ``aromatic infrared bands'' (AIBs).
    Even though specific molecules
    have not been assigned,
    the dominant {\it aromatic} nature
    of these bands is beyond doubt
    (e.g., see Peeters et al.\ 2004, Yang \& Li 2023).
    }
ubiquitously seen in a wide variety of astrophysical
regions, have long been attributed to the C--H and
C--C stretching and bending vibrational modes of
PAHs (L\'eger \& Puget 1984, Allamandola et al.\ 1985,
Candian et al.\ 2018, Li 2020).

Despite the remarkable success of the assignment
of the UIE bands to a cosmic mixture of PAHs of
different sizes and charging states
(e.g., see Allamandola et al.\ 1999, Li \& Draine 2001),
a long-standing criticism had remained
ever since the PAH model was first proposed
in the 1980s, that is, not a single specific PAH molecule
had been identified in the interstellar space
(e.g., see Kwok \& Zhang 2011).
However, this story has just changed in recent years.

Utilizing the 100\,m Green Bank Telescope (GBT),
McGuire et al.\ (2021) carried out radio observations
of the Taurus Molecular Cloud (TMC)
in the frequency range of 8 to 34$\GHz$.
The rotational transitions of
two specific PAH species---1-cyanonaphthalene
(1-CNN) and 2-cyanonaphthalene (2-CNN),
two isomers of cyanonaphthalene
(C$_{10}$H$_7$CN) in which a cyano (--CN)
group replaces one of the hydrogen atoms
of naphthalene (C$_{10}$H$_8$)---were detected 
in the dark molecular cloud TMC-1.
%
Similarly, also based on radio observations 
with GBT in the frequency range of
2--12 and 18--34$\GHz$,
Burkhardt et al.\ (2021) detected the rotational
transitions of indene (C$_9$H$_8$),
also in the TMC-1 cloud.
Cernicharo et al.\ (2021) also detected
the rotational lines of indene in TMC-1
in the frequency range of 31--50$\GHz$,
using the Yebes 40\,m radio telescope.
While 1- and 2-CNN are the {\it first} specific PAH
species ever identified in the interstellar medium
(ISM), indene, composed of both a five- and
six-membered ring, is the first {\it pure} PAH
molecule ever detected in space.
Cyanonaphthalenes are not pure PAH species,
instead, they are derivatives of naphthalene,
substituting a CN group for a hydrogen atom.
%
More recently, Sita et al.\ (2022) reported
the detection of 2-cyanoindene
(1H-indene-2-carbonitrile; 2-C$_9$H$_7$CN),
an isomer of cyanoindene
(which is a nitrile derivative of indene),
also in the TMC-1 cloud.
This detection, also made with 
the GBT at centimeter wavelengths,
provides the first direct measure of
the ratio of a cyano-substituted PAH molecule
to its pure hydrocarbon counterpart
in the same source.

The detection of indene, 2-cyanoindene,
1- and 2-CNN as well as benzonitrile
(C$_6$H$_5$CN; McGuire et al.\ 2018),
a single benzene ring with an attached CN group,
through radio observations
of their rotational spectra relies on the fact
that these molecules have a reasonably large
electric dipole moment. 
Although benzene (the parent molecule
of benzonitrile) and naphthalene (C$_{10}$H$_{8}$;
the parent molecule of 1- and 2-CNN)
lack a permanent dipole moment and
therefore have no pure rotational spectra,
the presence of a cyano group
dramatically increases the dipole moment
of their CN-substituted derivatives.
On the other hand, with an asymmetrical
structure, indene is a polar molecule
and has a permanent dipole moment
and this enables its detection through
rotational spectroscopy. 

However, neither cyanonaphthalenes nor indene
have been detected through their IR
vibrational signals. As a matter of fact,
IR identification of specific PAH molecules
has not yet been successful
even for just a single species.
Although the UIE bands are widely used as evidence
for the existence of PAHs in space,
they are mostly indicative of the general class of molecules.
Therefore, identifying which specific molecular species
are present through the UIE bands is difficult.
Nevertheless, as demonstrated in the {\it NASA/Ames
PAH Database} (see Boersma et al.\ 2014,
Bauschlicher et al.\ 2018, Mattioda et al.\ 2020),
a treasure library of a large variety
of specific neutral and cationic molecules
(and relatively fewer anions) of different
sizes and structures,
individual PAH molecules {\it do} exhibit certain
characteristic IR features.
The nondetection of individual PAH molecules
in the IR could be partly attributed to the lack of
knowledge of their exact IR spectra,
particularly in the far-IR.\footnote{%
  Although rather arbitrary, we divide
  the IR wavelength range into
  the near-IR ($\lambda$\,$\simali$1--5$\mum$),
  mid-IR ($\lambda$\,$\simali$5--15$\mum$),
  and far-IR $\lambda$\,$\simgt$\,15$\mum$).
  } 
Compared to the near- and mid-IR vibrational bands
which are more representative of functional groups,
the far-IR bands are sensitive to the skeletal characteristics
of a molecule, hence they contain specific fingerprint
information about the molecular identity of a PAH molecule.
However, without a prior knowledge of which specific
molecules are present in space, searching for the IR
signals of an individual PAH molecule is like finding
a needle in a haystack, since interstellar PAHs are
unlikely a single species, but a cosmic mixture of
a large number of species of different sizes
and charge states
(e.g., see Allamandola et al.\ 1999, Li \& Draine 2001).
The definite detection of individual, specific PAH
molecules such as cyanonaphthalenes and indene
made through their rotational transitions
(McGuire et al.\ 2021, Burkhardt et al.\ 2021),
therefore, offers us a valuable, effective guidance
to search for their IR signals in the ISM.

For an individual PAH molecule,
even if it is indeed present in space,
its abundance is not expected to be high.
Note that interstellar PAHs as a family lock up
$\simali$40--60 parts per million (ppm)
of the interstellar carbon
(relative to hydrogen; see Li \& Draine 2001).
If an individual molecule accounts for
$\simali$1\% of the PAH abundance,
its IR emission is expected to be weaker than
the UIE bands by a factor of $\simali$100.
While it may be challenging for
the {\it Infrared Space Observatory} (ISO)
and even the {\it Spitzer Space Telescope}
to detect such weak signals,
with the advent of
the {\it James Webb Space Telescope} (JWST),
this may become possible
due to its unprecedented sensitivity.\footnote{%
    Indeed, benzene has recently been identified
    by JWST in a protoplanetary disk
    around a low-mass star,
    based on the detection of three emission features
    around 14.85$\mum$ which are attributed to
    the Q branches of the fundamental and hot bending
    mode $\nu_4$ of benzene (Tabone et al.\ 2023).
    The 14.85$\mum$ $\nu_4$ bending band
    has previously been seen in {\it absorption}
    by the {\it Infrared Space Observatory} (ISO)
    in CRL~618, a protoplanetary nebula
    (Cernicharo et al.\ 2001).
   }
To facilitate JWST to search for the IR signals
of specific PAH molecules in space,
we perform a systematic exploration
of the theoretical IR emission spectra of various
specific PAH species expected in different
astrophysical environments.

In this work we model the vibrational excitation
of indene in a number of representative environments
and calculate its IR emission spectrum.
In a separate paper we will present 
the model IR emission spectra of
cyanonaphthalenes (Li et al.\ 2024).
Because of the energy balance,
the amount of power emitted in the IR
should equal to that absorbed
in the ultraviolet (UV).
Therefore, we first synthesize in \S\ref{sec:cabs} 
the UV absorption cross sections
of indene from the experimental data
available in the literature.
The UV absorption cross sections
determine how indene
absorbs starlight in the ISM.
In \S\ref{sec:cabs} we present 
the IR absorption cross sections
of indene which determine how it emits in the IR.
The vibrational excitation and de-excitation
processes of indene are discussed in 
\S\ref{sec:model}.
The model IR emission spectra of indene
are presented in \S\ref{sec:irem}
and discussed in \S\ref{sec:discussion}.
The major results are summarized in \S\ref{sec:summary}. 
%

\begin{figure}
\hspace{-12mm}
\includegraphics[height=8cm,width=10.8cm]{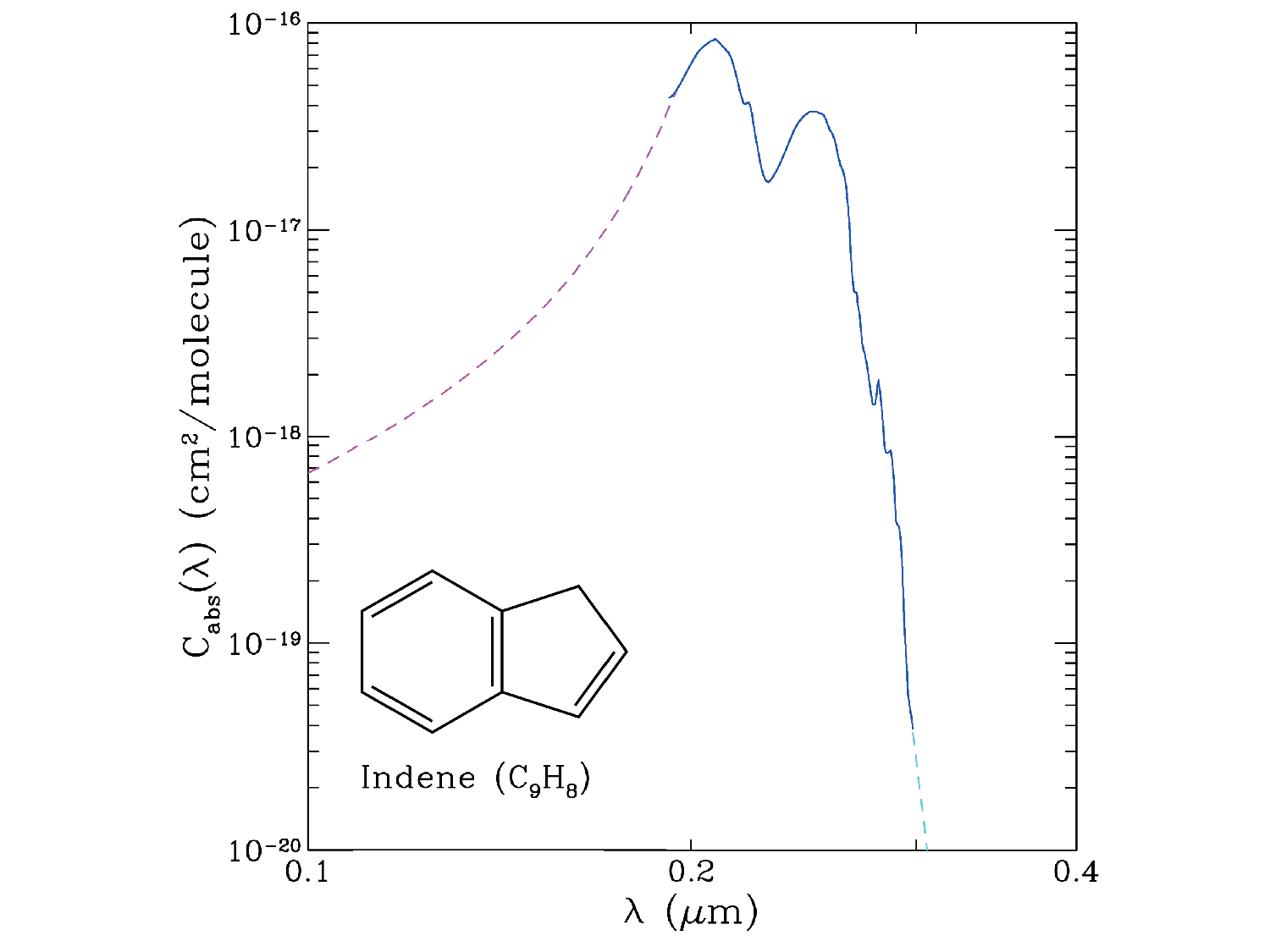}
\caption{
         \label{fig:cabs_uv}
         UV absorption cross sections (per molecule)
         of indene. Solid blue line: the experimental data
         of Perkampus (1992)
         at $0.19\mum <\lambda<0.30\mum$.
         Dashed magenta line at $\lambda<0.19\mum$
         and dashed cyan line at $\lambda>0.30\mum$
         are extrapolated from that of Perkampus (1992).
         Inserted is the chemical structure of indene.
         }
\end{figure}

\section{UV and IR Absorption Cross Sections}\label{sec:cabs}
The UV absorption has been measured for indene
over the wavelength range of
$0.19\mum <\lambda<0.30\mum$ (Perkampus 1992).
For the absorption at $\lambda<0.19\mum$
and $\lambda>0.30\mum$, 
we lack laboratory data
and therefore have to extrapolate from
the experimental data of Perkampus (1992)
at $0.19\mum <\lambda<0.30\mum$.
Figure~\ref{fig:cabs_uv} shows
the resulting UV absorption cross sections
of indene. It is apparent that indene absorbs
very little at $\lambda>0.30\mum$.
This is not unexpected as the absorption edge
occurs at shorter wavelengths for smaller PAHs
(see Appendex A2 of Li \& Draine 2001).
Admittedly, the extrapolation of the absorption 
at $\lambda<0.19\mum$ is somewhat simplified.
If additional absorption bands are present
at $\lambda<0.19\mum$, 
the UV absorption of indene would be higher
than that represented by the dashed magenta
line in Figure~\ref{fig:cabs_uv}.
However, we will show in \S\ref{sec:discussion},
with an enhanced UV absorption,
the IR emission spectrum of indene remains
essentially the same, except the overall
emission intensity is higher.

\begin{figure}
\centering
\includegraphics[height=8cm,width=8cm]{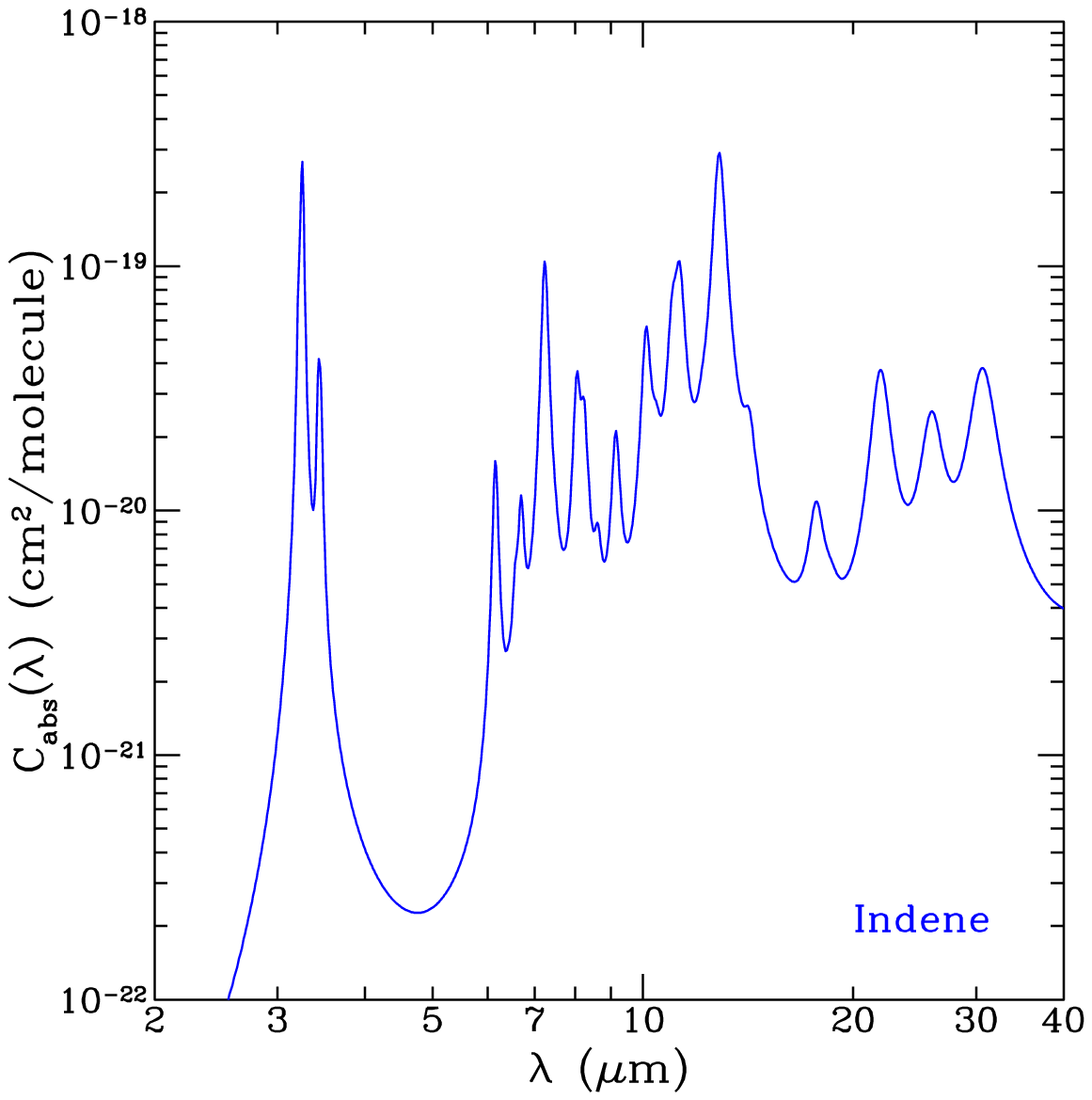}
\caption{
         \label{fig:cabs_ir}
         IR absorption cross sections (per molecule)
         of indene calculated from the vibrational frequences
         and intensities computed by the NASA/Ames
         Astrochemistry Group
         using B3LYP/4-31G.
         Each vibrational line is assigned
         a width of 30$\cm^{-1}$, consistent with
         the natural line width expected from
         a vibrationally excited PAH molecule 
         (see Allamandola et al.\ 1999).
         }
\end{figure}

In the IR, the vibrational frequencies and intensities
of indene have been computed by the NASA/Ames
Astrochemistry Group, using the B3LYP density
functional theory (DFT) in conjunction with
the 4-31G basis set. We take the computed
vibrational frequencies and intensities of indene
from the {\it NASA Ames PAH IR Spectroscopic Database}
(Boersma et al.\ 2014, Bauschlicher et al.\ 2018, 
Mattioda et al.\ 2020).
We represent each vibrational transition
by a Drude function,
characterized by the peak wavelength and intensity
of the vibrational transition.
For each transition, we assign a width of 30$\cm^{-1}$,
consistent with the natural line width
expected from a vibrationally excited PAH molecule 
(see Allamandola et al.\ 1999).
Figure~\ref{fig:cabs_ir} shows
the resulting IR absorption sections of indene.
We note that Klots (1995) had measured
the vibrational spectra of indene vapor 
using the {\it Fourier Transform IR Spectroscopy}
and Raman spectroscopy.
On the other hand,
Zilberg et al.\ (1996) had performed {\it ab initio}
calculations. 
The major absorption bands
seen in Figure~\ref{fig:cabs_ir} can also be found
in the experimental spectra of Klots (1995)
and the computational spectra of Zilberg et al.\ (1996).
However, neither Klots (1995) nor Zilberg et al.\ (1996)
reported the band intensities;
also, the vibrational frequencies were not
reported for all the modes.
In modeling the IR emission of indene,
a knowledge of the frequencies of all vibrational modes 
and the intensities of the major bands is required.

\section{Vibrational Excitation of
           Indene}\label{sec:model}
Indene has 17 atoms and 45 vibrational degrees of freedom. 
With such a small number of atoms,
its energy content (which is proportional to
the number of atoms) is often smaller than
the energy of a single stellar photon.  
Therefore, upon absorption of an UV stellar photon,
indene will undergo stochastic heating.
Following the absorption of an energetic photon, a PAH molecule
has three major competing decay channels to relax its energy: 
radiation,\footnote{%
  The emission process is dominated by fluorescence 
  (transitions between vibrational states of same 
  multiplicity) in the IR and part in the visible.
  Phosphorescence (transitions between vibrational states 
  of different multiplicity) is less important
  (see Li 2004). In this work we will thus only
  consider IR emission via fluorescence.
  }
photoionization, and photodissociation (see Li 2004).
In this work, we assume that the absorbed
starlight photon energy is exclusively lost
through radiative relaxation
(i.e., ignoring ionization and dissociation).
Such an assumption is a common practice
in modeling the IR emission of PAHs
(e.g., see Draine \& Li 2001, Draine et al.\ 2021).
%
At conditions with intense and hard UV photons,
indene may be ionized.
However, there are no experimental or computational
data on the photoionization and electron-recombination
of indene. This prevents a quantitative determination
of the ionization fraction of indene in the ISM.
Also, there lack data on the UV absorption
of ionized indene.\footnote{%
   Chalyavi et al.\ (2013) had measured
   the gas-phase electronic absorption 
   spectrum of indene cation in the visible
   and near-UV regions
   at $\simali$3000--6000$\Angstrom$,
   using resonance-enhanced photodissociation
   spectroscopy of He- and Ar-tagged ions.
   More recently, Chu et al.\ (2023) obtained
   the electronic spectrum of untagged indene
   cation in the $\simali$5500–-5800$\Angstrom$
   wavelength range, using sensitive cavity ring-down
   spectroscopy in a supersonically expanding planar
   plasma. However, neither Chalyavi et al.\ (2013)
   nor Chu et al.\ (2023) could measure the absorptivities
   which are required to determine how much power
   an indene cation would absorb when exposed to
   starlight heating. Also, as the interstellar radiation
   field is often far stronger at $\lambda<0.3\mum$
   than at $\lambda>0.3\mum$, the measured absorption
   spectra of Chalyavi et al.\ (2013) and Chu et al.\ (2023)
   do not have the wavelength coverage to properly
   determine the heating of cationic indene.
   }
This makes it difficult to model its vibrational
excitation and calculate its IR emission spectrum.
Since so far only the neutral forms of specific
PAH molecules have been detected in the ISM,
in this work we therefore limit ourselves to
{\it neutral} indene. 

To model the vibrational excitation and
radiative relaxation of indene, we take 
the ``exact-statistical'' method
developed by Draine \& Li (2001).
We characterize the state of indene
by its vibrational energy $E$, 
and group its energy levels into 
$(\Nb+1)$ ``bins'',
where the $j$-th bin ($j$\,=\,0,\,...,\,$\Nb$) 
is $[E_{j,\min},E_{j,\max})$, 
with representative energy 
$E_j$\,$\equiv$\,$(E_{j,\min}$+$E_{j,\max})$/2,
and width 
$\Delta E_j$\,$\equiv$\,$(E_{j,\max}$--$E_{j,\min})$.
Let $P_j$ be the probability of finding 
indene in bin $j$ with energy $E_j$.
The probability vector $P_j$ evolves 
according to
\begin{equation}
dP_i/dt = \sum_{j\neq i} \bT_{ij} P_j 
- \sum_{j\neq i} \bT_{ji}P_i ~~,~~ i\,=\,0,...,\Nb ~~,
\end{equation}
where the transition matrix element $\bT_{ij}$ is
the probability per unit time for indene in bin $j$ 
to make a transition to one of the levels in bin $i$. 
We solve the steady state equations
\begin{equation}\label{eq:steadystate}
\sum_{j\neq i} \bT_{ij} P_j
= \sum_{j\neq i} \bT_{ji}P_i ~~,~~ i\,=\,0,...,\Nb ~~
\end{equation}
to obtain the $\Nb$+1 elements of $P_j$,
and then calculate the resulting IR emission spectrum
(see eq.\,55 of Draine \& Li 2001).

In calculating the state-to-state transition rates
$\bT_{ji}$ for transitions $i$$\rightarrow$$j$,
we distinguish the excitation rates $\bT_{ul}$ 
(from $l$ to $u$, $l$\,$<$\,$u$) 
from the deexcitation rates $\bT_{lu}$ 
(from $u$ to $l$, $l$\,$<$\,$u$).
For a given starlight energy density $u_E$,
the rates for upward transitions $l$$\rightarrow$$u$ 
(i.e., the excitation rates)
are just the photon absorption rates:
\begin{equation}\label{eq:Tul}
\bT_{ul} \approx C_{\abs}(E)\,c\,u_E \Delta E_u/(E_u-E_l) ~~.
\end{equation}
The rates for downward transitions $u$$\rightarrow$$l$
(i.e., the deexcitation rates)
can be determined from the detailed balance analysis
of the Einstein $A$ coefficient:
\begin{equation}\label{eq:Tlu}
\bT_{lu} \approx \frac{8\pi}{h^3c^2} \frac{g_l}{g_u}
\frac{\Delta E_u}{E_u-E_l} E^3 \times C_{\abs}(E) 
\left[1+\frac{h^3c^3}{8\pi E^3}u_E\right] ~~,
\end{equation}
where $h$ is the Planck constant,
and the degeneracies $g_u$ and $g_l$ 
are the numbers of energy states 
in bins $u$ and $l$, respectively:
\begin{equation}
\label{eq:gj}
g_j \equiv N(E_{j,\max})-N(E_{j,\min})
\approx \left(dN/dE\right)_{E_j} \Delta E_j ~~,
\end{equation}
where $\left(dN/dE\right)_{E_j}$
is the vibrational density of states 
of indene at internal energy $E_j$.

For indene, the frequencies
and the oscillator strengths
of all its 45 vibrational modes
have already been computed
from DFT/B3LYP in conjunction
with the 4-31G basis set 
(see Bauschlicher et al.\ 2018)
and made available
in the NASA/Ames PAH Database.
We adopt the Beyer-Swinehart numerical algorithm
(Beyer \& Swinehart 1973, Stein \& Rabinovitch 1973)
to calculate from these vibrational frequencies
the vibrational density of states 
$\left(dN/dE\right)_{E_j}$
and therefore the degeneracies $g_j$ 
for each vibrational energy bin.
As mentioned in \S\ref{sec:cabs},
we obtain the IR absorption
cross section $C_{\rm abs}(E)$
by summing up all the vibrational transitions
with each approximated as a Drude profile.

For a given astrophysical environment
characterized by its starlight energy density $u_E$,
we first calculate the excitation rates $\bT_{ul}$ 
(from $l$ to $u$, $l$\,$<$\,$u$)
according to eq.\,\ref{eq:Tul},
using the UV absorption cross setions
described in \S\ref{sec:cabs}.
We then calculate the deexcitation rates
$\bT_{lu}$ (from $u$ to $l$, $u$\,$>$\,$l$)
according to eq.\,\ref{eq:Tlu},
using the degeneracy $g_j$ and
the IR absorption cross section $C_{\rm abs}(E)$
derived from the DFT-computed vibrational
frequencies and intensities.
With the state-to-state 
transition rates $\bT_{ji}$ determined,
we solve the steady-state probability 
evolution equation (see eq.\,\ref{eq:steadystate})
to obtain the steady-state energy probability 
distribution $P_j$ 
and finally calculate 
the resulting IR emission spectrum,
according to eq.\,55 of Draine \& Li (2001).
For computational convenience, 
we consider 500 energy bins
(i.e., $M=500$).
%

\begin{figure}
\centering
\includegraphics[height=8cm,width=8cm]{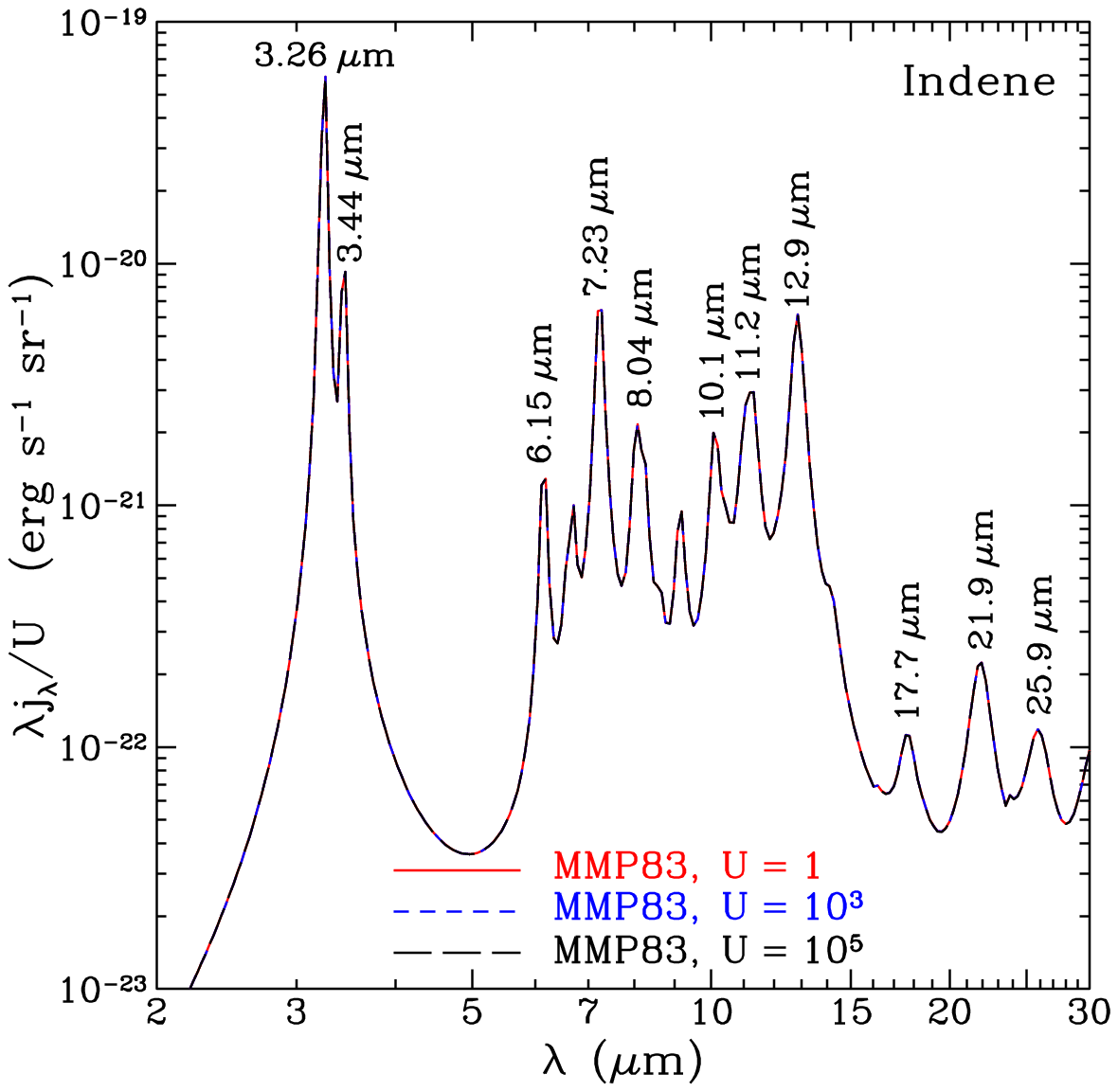}
\caption{
         \label{fig:mmp83}
         IR emission spectra of indene
         illuminated by the MMP83 radiation field
         with different intensities
         (red solid line: $U$\,=\,1;
         blue shot-dashed line: $U$\,=\,1,000;
         black long-dashed line: $U$\,=\,10$^5$).
         The C--H streches at 3.26 and 3.44$\mum$
         and the far-IR skeleton bending bands
         at $\simali$17.7, 21.9, and 25.9$\mum$
         may be used to probe the presence of
         indene in the ISM. 
         }
\end{figure}

\section{Model Emission Spectra}\label{sec:irem}
We first consider indene in the diffuse ISM 
excited by the solar neighborhood interstellar radiation
field of Mathis et al.\ (1983; hereafter MMP83).
As shown in Figure~\ref{fig:mmp83},
the IR emission spectrum of indene exhibits
a prominent C--H stretching band
at 3.26$\mum$ attributed to the six-membered
benzene ring, as well as a satellite band at 3.44$\mum$
arising from the C--H stretches of the five-membered ring.
Also prominent are two broad complexes at
$\simali$6--9$\mum$ and $\simali$10--14$\mum$
which are respectively composed of a number of 
C--C stretching bands at
$\simali$6.15, 7.23, 8.04 and 9.15$\mum$
and C--H out-of-plane bending bands
at $\simali$10.1, 11.2, and 12.9$\mum$.
In addition, indene also emits three weak
bands at $\simali$17.7, 21.9, and 25.9$\mum$
attributed to its skeleton vibrations.

To examine how the IR emission of indene
varies with the starlight {\it intensity}, 
we calculate the vibrational emission spectra
of indene excited by starlight {\it intenser}
than the MMP83 radiation field.
For illustration, we show in Figure~\ref{fig:mmp83}
the IR emission spectra of indene
exposed to MMP83-type radiation fields,
but enhanced by a factor of $U$\,=\,1000 and 10$^5$
($U$\,=\
l,1 corresponds to the MMP83 radiation field).
As demonstrated in Figure~\ref{fig:mmp83}, 
when scaled by $U$, the IR emission spectra
are identical for different starlight intensities $U$.
This is not unexpected: in the stochastic heating regime,
the IR emission spectral shape does not
vary with the starlight intensity
(see Figure~13 of Li \& Draine 2001,
Figure~13 of Draine \& Li 2007,
and Figure~10 of Draine et al.\ 2021).

We also examine how the IR emission of indene
varies with the starlight {\it spectrum}.
The spectral shape of the exciting starlight
characterizes the ``hardness'' of the radiation field.
We consider indene exposed to stars with different
effective temperatures:
$\Teff$\,=\,40,000, 22,000, 8,000$\K$.
The starlight spectra are approximated by
the Kurucz model atmospheric spectra.

\begin{figure*}
\begin{center}
\hspace{-1cm}
\begin{minipage}[t]{0.4\textwidth}
\resizebox{8.0cm}{7.5cm}{\includegraphics[clip]{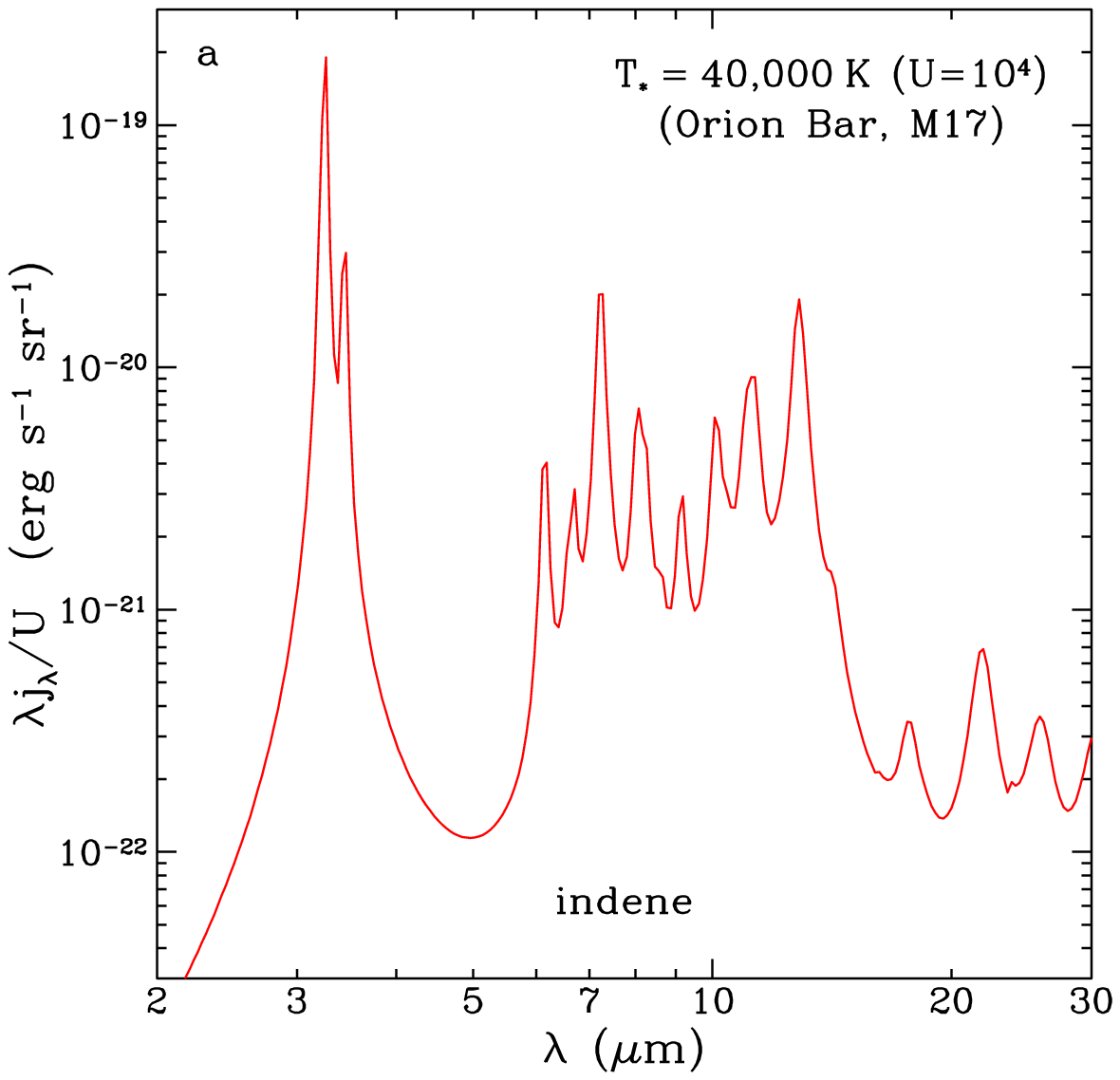}}\vspace{0.1cm}
\resizebox{8.0cm}{7.5cm}{\includegraphics[clip]{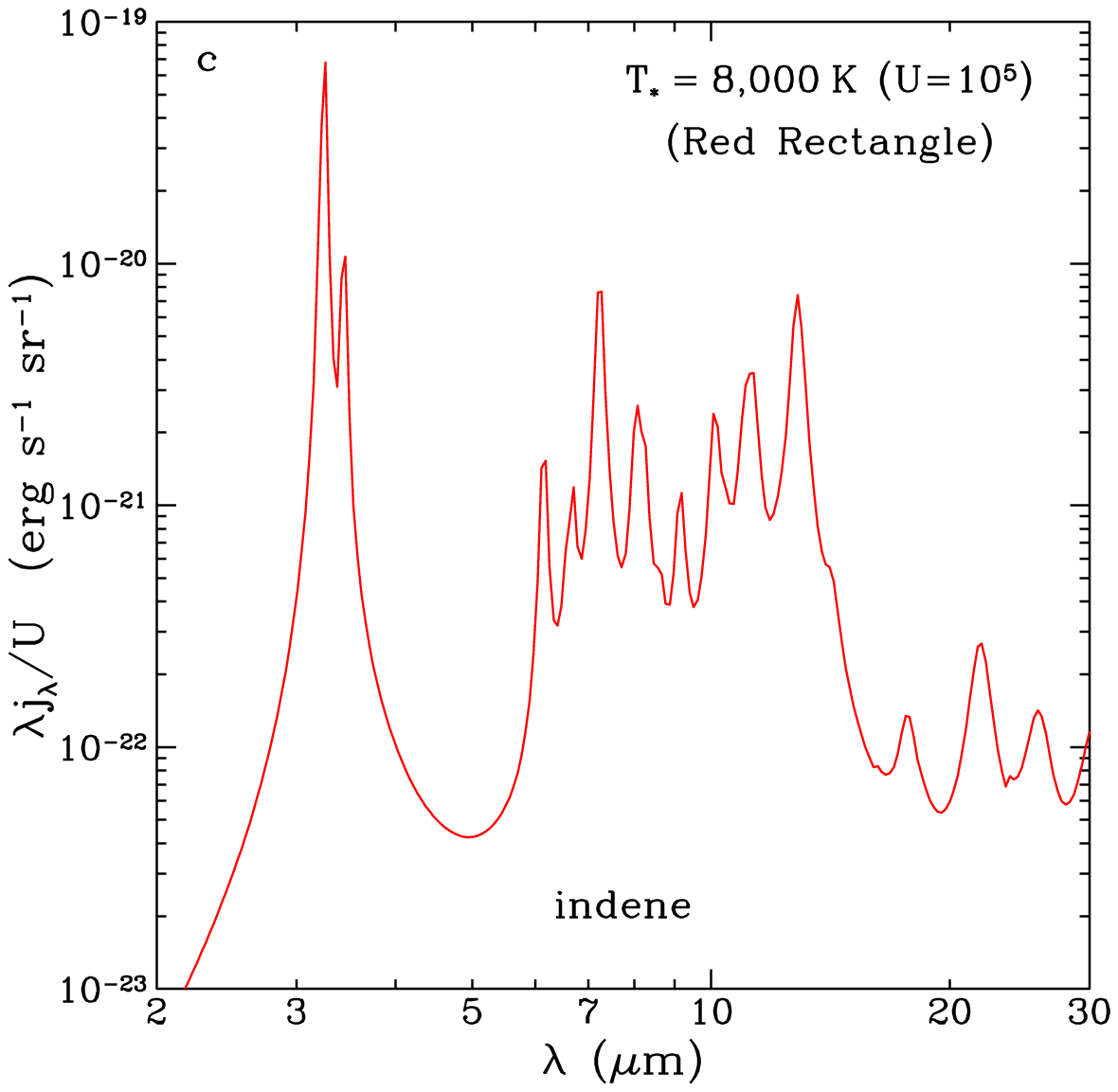}}\vspace{-0.5cm}
\end{minipage}
\hspace{2cm}
\begin{minipage}[t]{0.4\textwidth}
\resizebox{8.0cm}{7.5cm}{\includegraphics[clip]{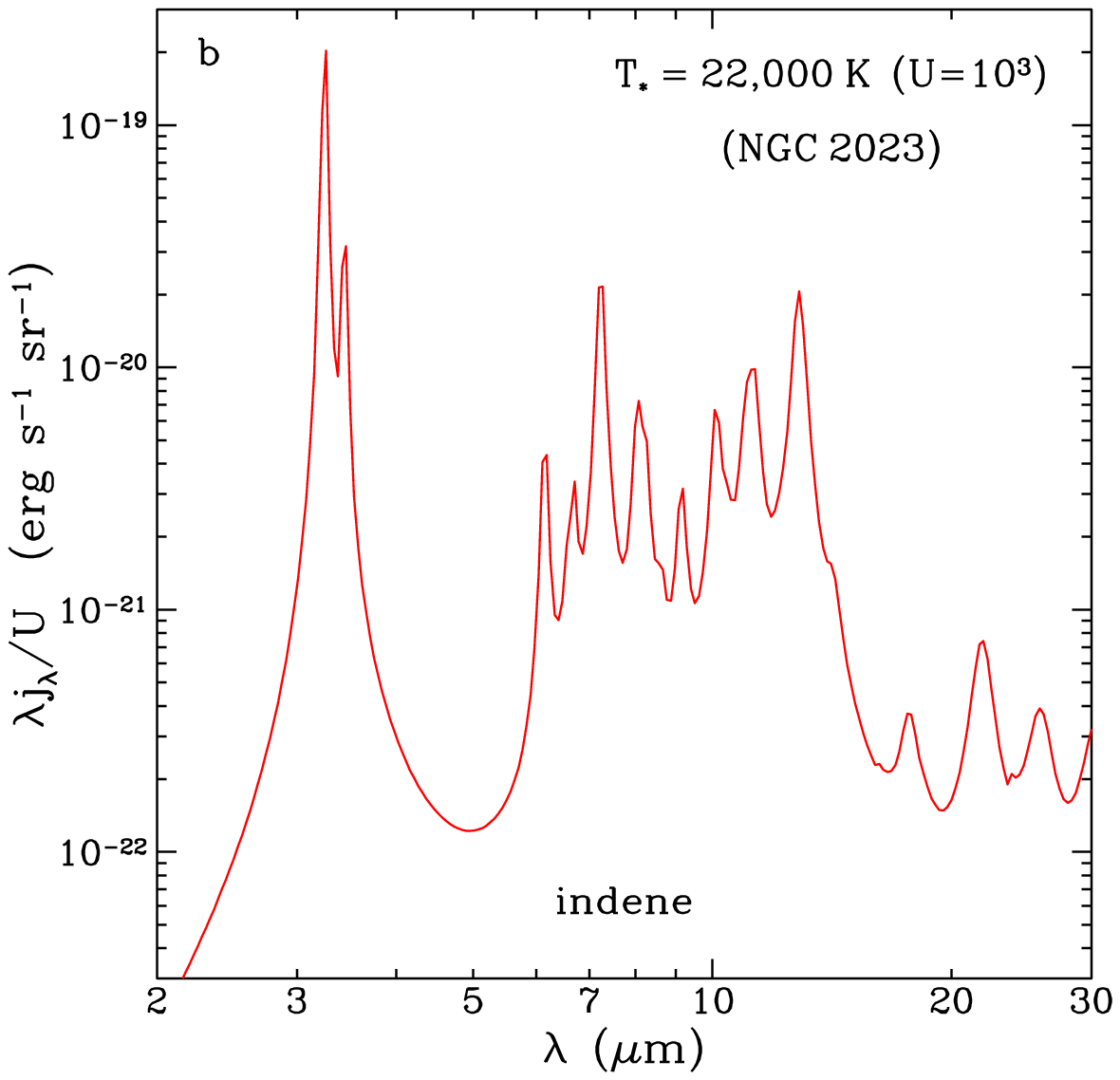}}\vspace{0.1cm}
\resizebox{8.0cm}{7.5cm}{\includegraphics[clip]{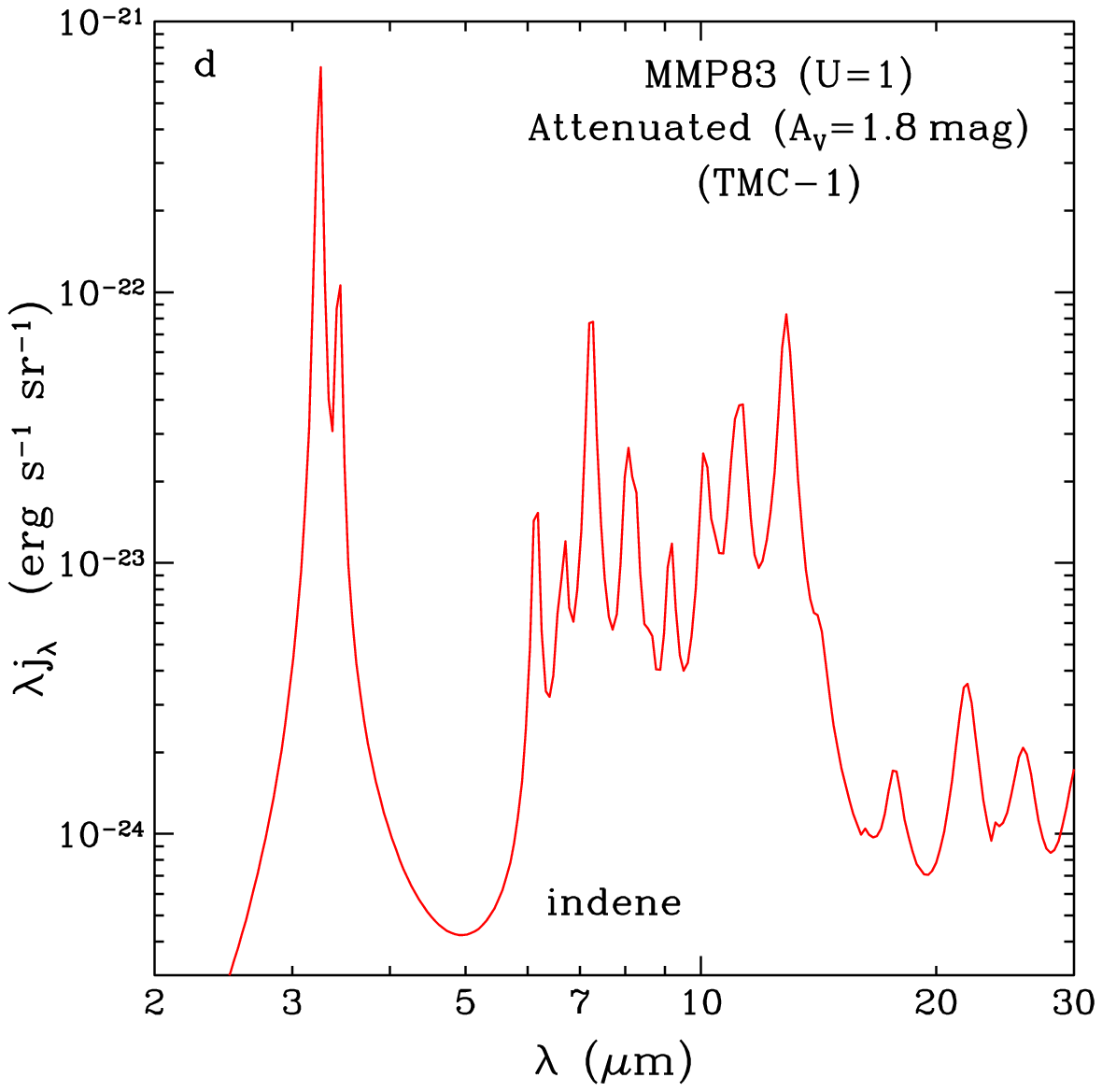}}\vspace{-0.5cm}
\end{minipage}
\end{center}
\caption{
         \label{fig:Teff}
         IR emission spectra of indene
         illuminated by radiation fields
         of different starlight spectra:
         $\Teff$\,=\,40,000$\K$
         and $U$\,=\,10$^4$ (top left panel),
         $\Teff$\,=\,22,000$\K$
         and $U$\,=\,10$^4$ (top right panel),
         $\Teff$\,=\,8,000$\K$
         and $U$\,=\,10$^5$ (bottom left panel),
         and MMP83 radiation field
         attenuated by a visual extinction
         of $A_V=1.8\magni$ (bottom right panel).
         }
\end{figure*}

Figure~\ref{fig:Teff}a shows the IR emission of indene
illuminated by stars of $\Teff$\,=\,40,000$\K$
(like O6V stars) with an intensity of $U$\,=\,10$^4$,
with $U$ defined as
\vspace{-3mm}
\begin{equation}
U = \frac{\int_{1\mu {\rm m}}^{912{\rm \Angstrom}}
               4\pi J_\star(\lambda,\Tstar)\,d\lambda}
       {\int_{1\mu {\rm m}}^{912{\rm \Angstrom}}
               4\pi J_{\rm ISRF}(\lambda)\,d\lambda} ~~,
\end{equation}
where $J_{\rm ISRF}(\lambda)$ is 
the MMP83 radiation intensity, and
$J_\star(\lambda, \Tstar)$ is the intensity
of starlight approximated by the Kurucz model
atmospheric spectrum.
Such a starlight spectrum and intensity
resemble that of the Orion Bar
photodissociation region
and the M17 star-forming region.
A comparison of Figure~\ref{fig:Teff}a 
with  Figure~\ref{fig:mmp83} reveals
that the IR emission spectra are almost
identical for indene excited by the MMP83 field
and by stars of $\Teff$\,=\,40,000$\K$,
except the emissivity level is higher for
the latter since the mean absorbed photon
energy is higher for the latter.

Figure~\ref{fig:Teff}b and Figure~\ref{fig:Teff}c
respectively show the IR emission of indene
illuminated by stars of $\Teff$\,=\,22,000$\K$
(like B1.5V stars) with an intensity of $U$\,=\,10$^3$
and of $\Teff$\,=\,8,000$\K$ (like A5V stars)
with an intensity of $U$\,=\,10$^5$.
While the former applies to the reflection
nebula NGC\,2023, the latter is like
the Red Rectangle protoplanetary nebula.
When scaled by $U$, the IR emission spectra
for both $\Teff$\,=\,22,000$\K$
and $\Teff$\,=\,8,000$\K$ are essentially
identical to that for $\Teff$\,=\,40,000$\K$.

As indene has been detected in the TMC-1 molecular cloud
through its rotational lines, we calculate the IR emission
expected for indene in TMC-1, which is externally
illuminated by the general interstellar radiation field
but attenuated by dust extinction with a visual extinction
of $\simali$3.6$\magni$ (Whittet et al.\ 2004).
We approximate the starlight by 
the MMP83 interstellar radiation field
attenuated by $\exp\{-\left(A_V/1.086\right)
\times\left(A_\lambda/A_V\right)\}$,
where $A_V$\,=\,1.8$\magni$ is the visual
extinction, $A_\lambda$ is the extinction at
wavelength $\lambda$, and $A_\lambda/A_V$,
the wavelength-dependence of extinction,
is taken to be that of
the Galactic average extinction curve
of $R_V=3.1$, where $R_V$ is the total-to-selective
extinction ratio (see Cardelli et al.\ 1989).
Figure~\ref{fig:Teff}d shows the IR emission
of indene calculated for the TMC-1 cloud.
Compared to that excited by the MMP83
radiation field (see Figure~\ref{fig:mmp83})
and stars of $\Teff$\,=\,40,000, 22,000, 8,000$\K$,
the overall IR emission spectrum of indene
expected in the TMC-1 cloud is closely similar.
%

\section{Discussion}\label{sec:discussion}
We have shown in \S\ref{sec:irem} that the spectral
shape of the IR emission of indene does not vary
with the spectral shape of the illuminating starlight
radiation field (see Figures~\ref{fig:mmp83}, \ref{fig:Teff}a--d).
This is because indene only has 45 vibrational degrees
of freedom and its heat capacity is so small that even
photons of a couple of electron volts can excite indene
sufficiently to emit in its vibrational modes.
In the TMC-1 molecular cloud, the mean energy of
photons absorbed by indene is $\hnuabs\approx3.9\eV$,
considerably lower than that in the diffuse ISM
($\hnuabs\approx5.9\eV$) and in regions
illuminated by stars with 
$\Teff$\,=\,40,000$\K$ ($\hnuabs\approx6.2\eV$),
22,000$\K$ ($\hnuabs\approx6.1\eV$), and
8,000$\K$ ($\hnuabs\approx5.7\eV$).
This demonstrates that the hardness of the starlight
does not affect the IR emission spectral shape of indene.

\begin{figure*}
\begin{center}
\hspace{-1cm}
\begin{minipage}[t]{0.4\textwidth}
\resizebox{8.5cm}{7.5cm}{\includegraphics[clip]{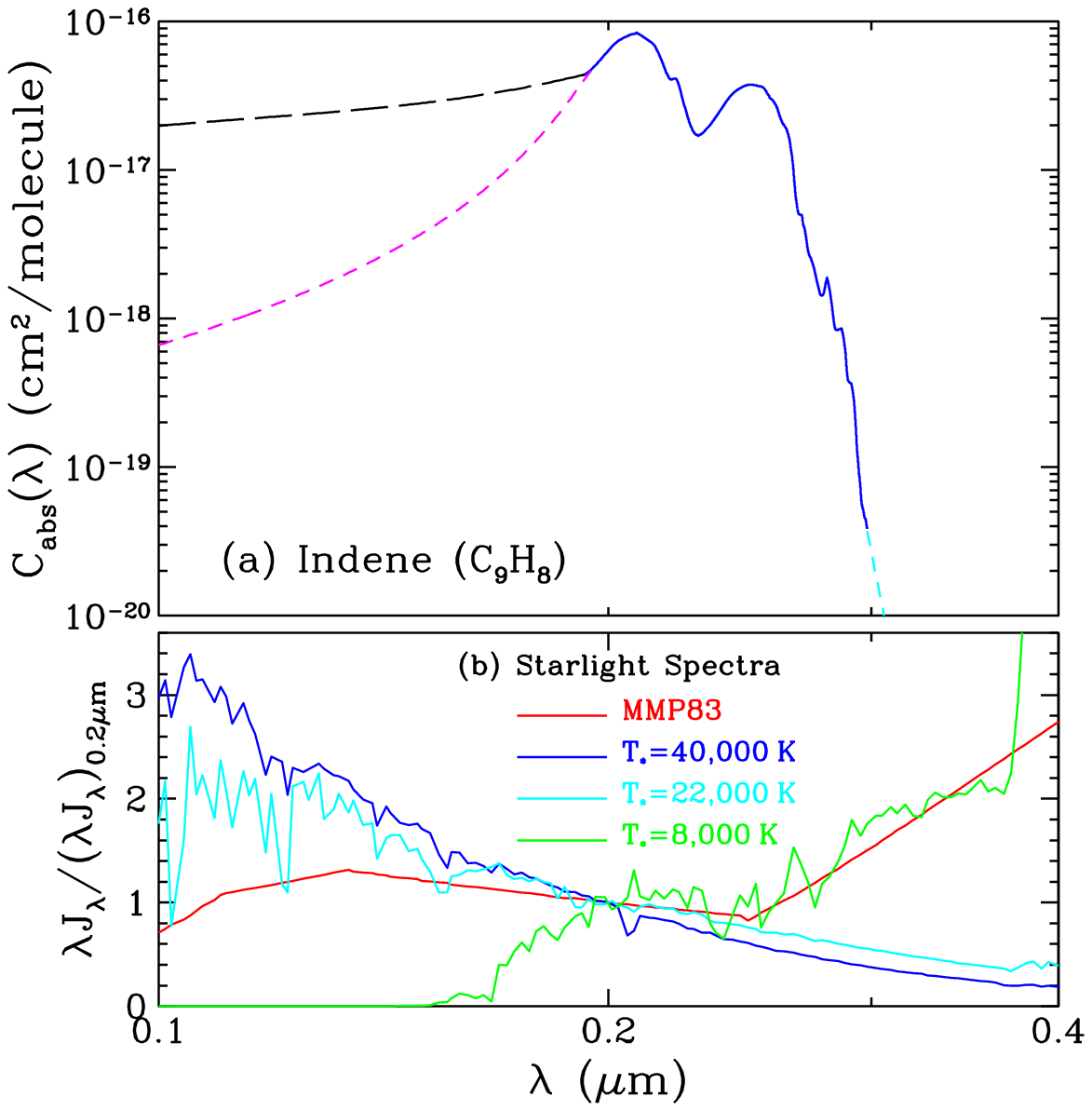}}\vspace{-0.5cm}
\end{minipage}
\hspace{2cm}
\begin{minipage}[t]{0.4\textwidth}
\resizebox{8.5cm}{7.5cm}{\includegraphics[clip]{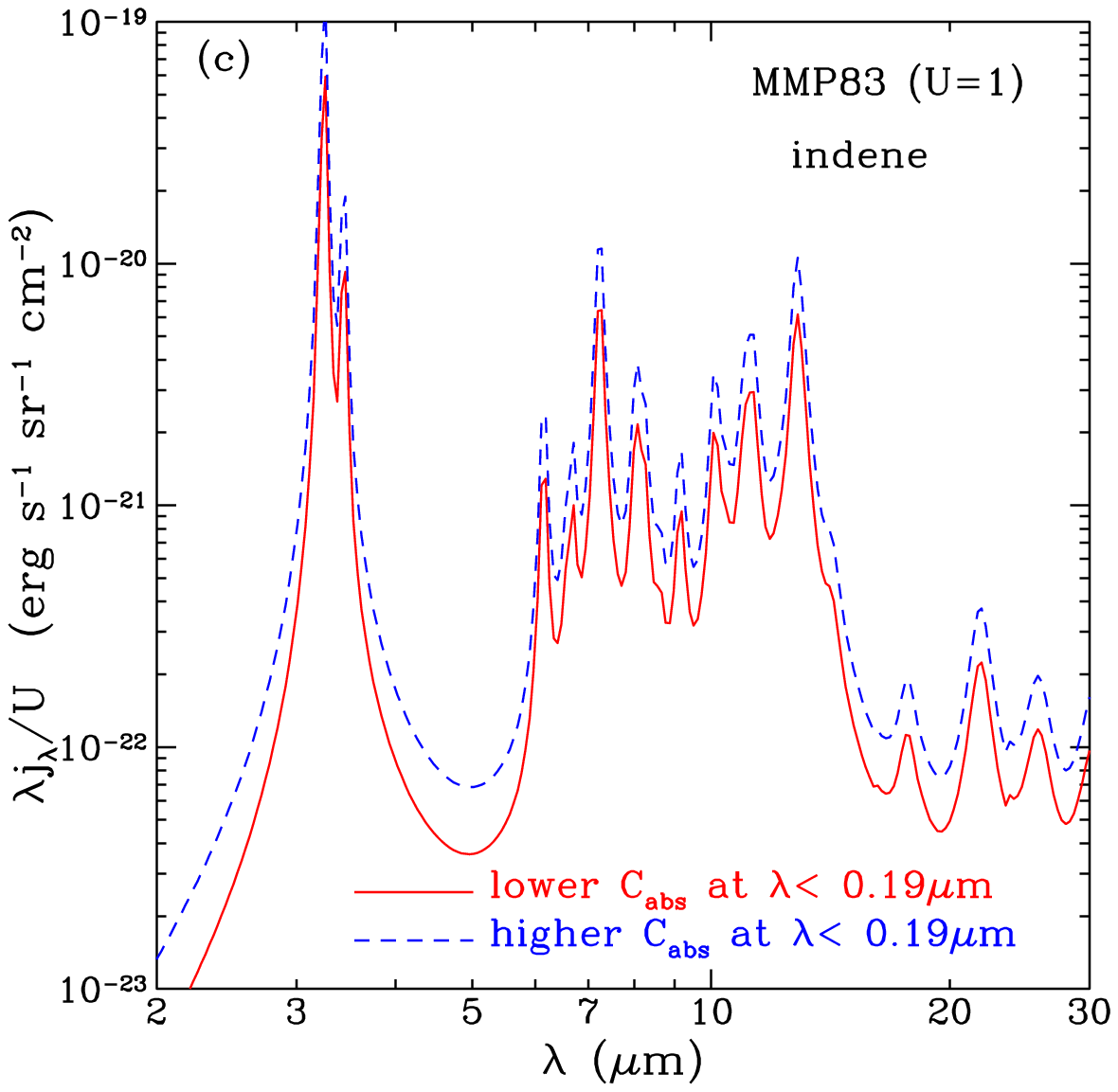}}\vspace{-0.5cm}
\end{minipage}
\end{center}
\caption{
         \label{fig:high_uvabs}
         Top left panel (a): Comparison of the UV absorption of
         indene with an elevated value at $\lambda<0.19\mum$
         (``higher $C_{\rm abs}$''; black long-dashed line)
         and that with the absorption at $\lambda<0.19\mum$
         extrapolated from that at $\lambda>0.19\mum$
         (``lower $C_{\rm abs}$''; magenta dashed line).
         The UV absorption at $\lambda>0.19\mum$
         for both cases is the same as Figure~\ref{fig:cabs_uv}.
         Bottom left panel (b): Comparison of
         the MMP83 interstellar radiation field (red)
         with the radiation fields represented by
         different stellar effective temperatures
         ($\Teff=40,000\K$: blue;
         $\Teff=22,000\K$: cyan; and
         $\Teff=8,000\K$: green).
         All the radiation intensities are normalized
         at $\lambda=0.2\mum$.
         Right  panel (c): Comparison of the model
         IR emission spectrum of indene in the diffuse
         ISM calculated with an elevated UV absorption
         at $\lambda<0.19\mum$ (dashed blue line)
         and that with the extrapolated UV absorption
         at $\lambda<0.19\mum$ (solid red line).
         }
\end{figure*}

We note that the model emission spectra presented
in \S\ref{sec:irem} were obtained by approximating
the UV absorption of indene at $\lambda<0.19\mum$
as a simple extrapolation from that at
$\lambda>0.19\mum$ (see Figure~\ref{fig:cabs_uv}).
If indene has additional absorption bands
at $\lambda<0.19\mum$, 
such an extrapolation would underestimate
the UV absorption of indene, and by implication,
underestimate the power emitted by indene in the IR.
To quantitatively examine this,
as shown in Figure~\ref{fig:high_uvabs}a,
we arbitrarily elevate its absorption cross
section at $\lambda<0.19\mum$ to a level
substantially exceeding the extrapolated one.
We also show in Figure~\ref{fig:high_uvabs}b
the starlight radiation fields
for the diffuse ISM (MMP83),
the Orion Bar or M17 ($\Teff=40,000\K$), 
NGC\,2023 ($\Teff=22,000\K$), and
the Red Rectangle ($\Teff=8,000\K$).
Figures~\ref{fig:high_uvabs}a and \ref{fig:high_uvabs}b
clearly show that, in the wavelength range 
where the starlight radiation fields
(except that of $\Teff=8,000\K$) are strong,
the adopted UV absorption at $\lambda<0.19\mum$ 
would substantially affect the heating of indene.
However, as shown in Figure~\ref{fig:high_uvabs}c,
the model IR emission spectrum
calculated with such an elevated
UV absorption 
for the diffuse ISM is essentially the same
as that with a lower, extrapolated UV absorption,
except that the overall emissivity is higher
by $\simali$70\%.
Note that, with such an elevated UV absorption
at $\lambda<0.19\mum$, the mean energy of
photons absorbed by indene in the diffuse ISM
is now $\hnuabs\approx7.1\eV$,
exceeding that with a lower UV absorption
($\hnuabs\approx5.9\eV$)
by $\simali$20\%.
This implies that the IR emission
cannot be simply scaled by $\hnuabs$.
This also implies that, in modeling
the IR emssion of small PAH molecules,
the assumption of a {\it single} photon
energy for their vibrational energy contents
is somewhat simplified.
To properly predict their IR emission,
we should consider the absorption
over the entire starlight spectrum.

The fact that the IR emission spectral shape of indene
remains essentially invariant both with the intensity
and with the hardness of the starlight radiation field
makes the identification of indene through its vibrational
emission spectrum simpler. However, the broad emission
complex at $\simali$6--14$\mum$ occurs at the same
wavelength range of the 6.2, 7.7, 8.6, 11.3 and 12.7$\mum$
UIE bands. It is therefore difficult to search for the IR
signals of indene through the $\simali$6--14$\mum$
emission complex. In contrast, the C--H stretching bands
at 3.26 and 3.44$\mum$ could potentially allow one to
identify indene in space.
While the {\it Infrared Spectrograph} (IRS)
on {\it Spitzer} only operates longward of
$\simali$5.2$\mum$ and does not
cover the characteristic 3.26 and 3.44$\mum$ bands
of indene, the {\it Short Wavelength Spectrometer} (SWS)
on board ISO and the {\it Infrared Camera} (IRC) on board
{\it AKARI} lack the sensitivity to detect weak signals
from specific PAH molecules such as indene.
The {\it Near InfraRed Spectrograph} (NIRSpec)
on JWST spans the wavelength range
of the characteristic C--H bands of indene.
This, combined with its unprecedented high sensitivity,
JWST could potentially place the detection of the C--H
bands of indene on firm ground.

\begin{figure}
\centering
\includegraphics[height=8cm,width=8cm]{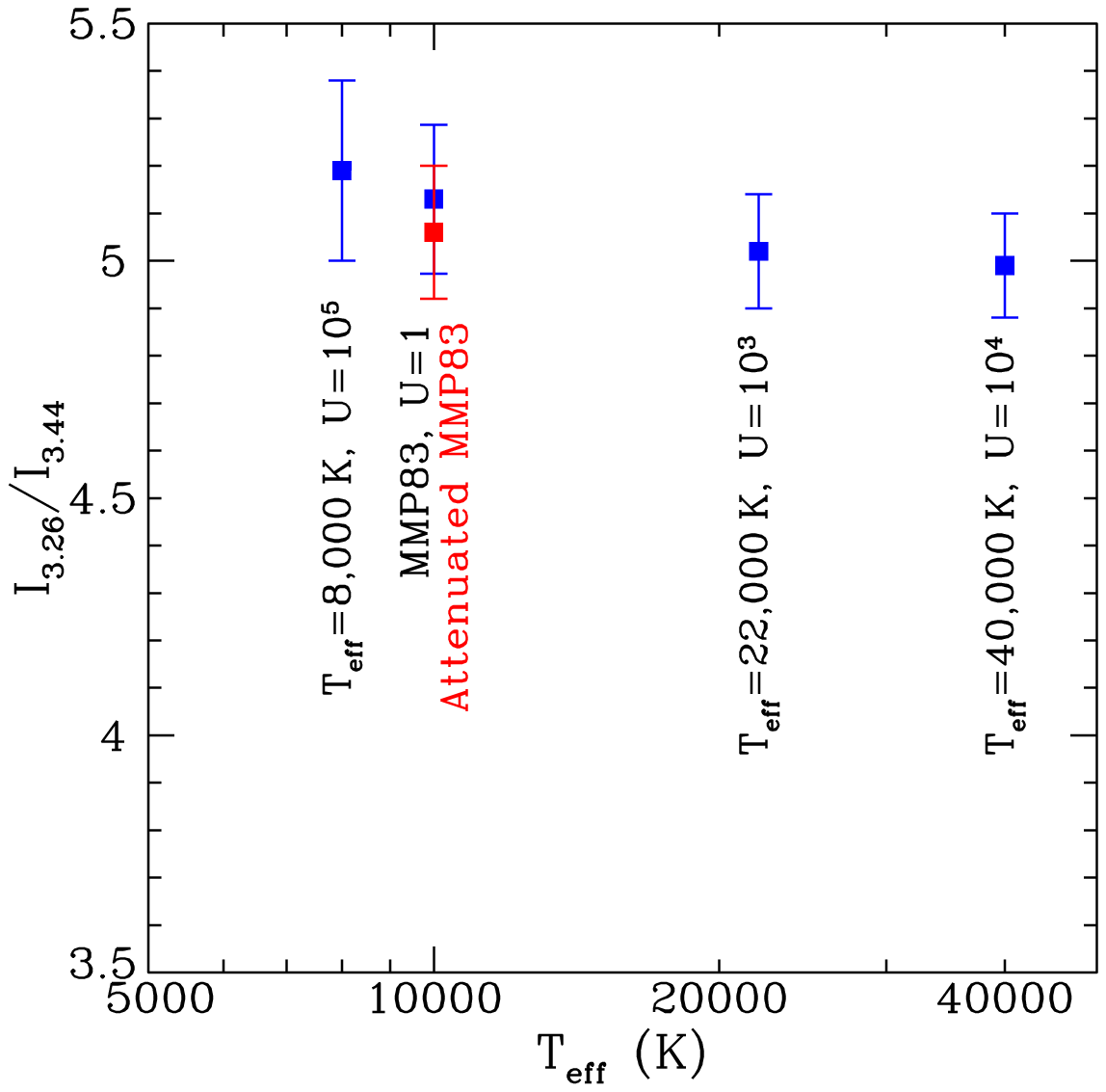}
\caption{
             \label{fig:BandRatio}
             Model-predicted 3.26$\mum$/3.44$\mum$
             band ratio of indene
             as a function of the effective temperature of
             the exciting star. The MMP83 interstellar radiation
             field is approximated by an effective temperatue
             of $\Teff=10,000\K$. Nearly a century ago,
             Eddington (1926) represented the interstellar
             radiation field in the solar vicinity by a black body
             of $10^4\K$ diluted by a factor of $10^{-14}$.
             }
\end{figure}

It should be noted that the characteristic 3.26 and
3.44$\mum$ bands of indene are close in wavelengths
to the 3.3$\mum$ UIE band and its satellite band
at 3.4$\mum$, respectively.
The 3.4$\mum$ band is commonly attributed to
the aliphatic C--H stretch of PAHs with aliphatic
sidegroups (see Yang et al.\ 2017) 
or superhydrogenated PAHs (see Yang et al.\ 2020).
With a spectral resolution of $\simali$1000,
the JWST/NIRSpec {\it Multi-Object Spectroscopy}
may be capable of distinguishing these bands.
Moreover, as illustrated in Figure~\ref{fig:BandRatio},
the model-predicted 3.26$\mum$/3.44$\mum$
band ratio ($I_{3.26}/I_{3.44}$) is essentially constant
and insensitive to the physical conditions.
In contrast, the 3.3$\mum$/3.4$\mum$ band ratio 
shows a strong dependence on the hardness
of the exciting radiation field, as illustrated 
in Figure~15 of Yang \& Li (2003).
Furthermore, the far-IR bands of indene
at $\simali$17.7, 21.9, and 25.9$\mum$
(see Figures~\ref{fig:mmp83} and \ref{fig:Teff})
can also be used to probe its presence in the ISM 
as the far-IR bands arise from the bending
of the whole PAH skeleton 
(mostly out-of-plane) and are thus intrinsically 
related to the molecular structure
(see Zhang et al.\ 2010).

Let $\NH$ be the hydrogen column density
along the line of sight to an astrophysical region. 
Let $\left[{\rm C/H}\right]_{\rm ind}$
be the number of carbon atoms (per H nucleon)
locked up in indene.
The intensity of the IR emission
(erg$\s^{-1}\cm^{-3}\sr^{-1}$)
expected from indene would be
\begin{equation}
I_\lambda = j_\lambda\times\left(\frac{\NH}{9}\right)\times\Cind  ~~,
\end{equation}
where the denominator ``9'' accounts for the fact
that an indene molecule has nine carbon atoms.
The column density of indene is simply
\begin{equation}
  N_{\rm ind} = \left(\frac{\NH}{9}\right)\times\Cind  ~~.
\end{equation}
Therefore, by comparing the observed intensity
$I_\lambda^{\rm obs}$ with the model emissivity
$j_\lambda$ (e.g., see Figure~\ref{fig:Teff}a--d),
one can derive $N_{\rm ind}$. If $\NH$ is known,
then one can determine how much carbon
(per H nucleon) is locked up in indene.
On the other hand, if $N_{\rm ind}$ is known
from the measurements of its rotational lines,
one can predict the IR emission intensity
$I_\lambda$ by multiplying the model emissivity
shown in Figure~\ref{fig:Teff}a--d with $N_{\rm ind}$.

Finally, we admit that how indene forms and survives
in the ISM is still puzzling.
It is generally believed that only PAHs with more than
20 carbon atoms may survive in the ISM (see Tielens 2008).
However, Iida et al.\ (2022) recently found that
small carbon clusters with as few as nine carbon atoms
can be efficiently stablized with the aid of the so-called
``Recurrent Fluorescence''
(also known as Poincar\'e fluorescence,
see L\'eger et al.\ 1988),
a radiative relaxation channel
in which optical photons are emitted
from thermally populated electronically excited states.
Regarding the detection of indene in TMC-1,
Mat\'e et al.\ (2023) argued that volatile molecules
like indene should freeze to a large extent on the
ice mantles of dust grains in TMC-1, and therefore
the detection of gas-phase indene may require
the cycling of this molecule between the gas
and the solid, probably through balancing the freezing
and accretion of indene on solid grains by cosmic ray
sputtering (e.g., see Dartois et al.\ 2022).
On the other hand, current models, including
gas-phase and grain surface reactions,
underpredict the observed indene abundances
in TMC-1 by three orders of magnitude
(Burkhardt et al.\ 2021).
By exploiting crossed molecular beam experiments,
Doddipatla et al.\ (2021) suggested an unusual pathway
leading to the formation of indene
via a barrierless bimolecular reaction involving
the simplest organic radical---methylidyne (CH)---and
styrene (C$_6$H$_5$C$_2$H$_3$) through the hitherto
elusive methylidyne addition–cyclization–aromatization
(MACA) mechanism. While this could be a viable route
in dense clouds, it remains unclear how indene could
form in the diffuse ISM.

\section{Summary}\label{sec:summary}
We have modeled the vibrational excitation of
indene, the first pure, specific PAH species
ever identified in space, and have calculated its
IR emission spectra for a number of representative
astrophysical regions, from the diffuse ISM to regions
illuminated by stars of different effective temperatures.
Also calculated is the TMC-1 dark molecular cloud
where indene was detected through its rotational lines.
It is found that the IR emission spectral shape of indene
remains essentially invariant with environments.
With the most prominent bands occuring at
$\simali$3.26 and 3.44$\mum$,
indene may be detectable by the NIRSpec instrument
on board JWST.

\section*{Acknowledgements}
We thank the anonymous referees
for helpful comments and suggestions.
We thank B.M.~Broderick, B.T.~Draine,
B.A.~McGuire, and E.F.~van Dishoeck
for stimulating discussions.
KJL and TTF are supported by
the National Key R\&D Program of China
under No.\,2017YFA0402600,
and the NSFC grants 11890692,
12133008, and 12221003,
as well as CMS-CSST-2021-A04.
XJY is supported in part by
NSFC~12333005 and 12122302
and CMS-CSST-2021-A09.

\section*{Data Availability}
The data underlying this article will be shared
on reasonable request to the corresponding authors.


\end{document}